\documentclass{PoS}

\usepackage{epsfig,multicol,multirow,cite}
\usepackage{amsmath,subfigure,latexsym,amssymb}

\newcommand{\be}{\begin{equation}}
\newcommand{\ee}{\end{equation}}
\newcommand{\nn}{\nonumber}
\newcommand{\bea}{\begin{eqnarray}}
\newcommand{\eea}{\end{eqnarray}} 

\newcommand{\la}{\langle}
\newcommand{\ra}{\rangle}

\newcommand{\Z}{\mathbb{Z}}
\newcommand{\R}{{\kern+.25em\sf{R}\kern-.78em\sf{I} \kern+.78em\kern-.25em}}
\newcommand{\RR}{{\kern+.25em\sf{R}\kern-.6em\sf{I} \kern+.6em\kern-.25em}}
\newcommand{\N}{{\kern+.25em\sf{N}\kern-.78em\sf{I} \kern+.78em\kern-.25em}}
\newcommand{\C}{{\kern+.25em\sf{C}\kern-.50em\sf{I} \kern+.50em\kern-.25em}}

\newcommand{\vp}{\varphi}

\title{Extracting Physics from Topologically Frozen Markov Chains}

\ShortTitle{Topologically Frozen Markov Chains}

\author{\speaker{Urs Gerber}, Irais Bautista, Wolfgang Bietenholz,
H\'{e}ctor Mej\'{\i}a-D\'{\i}az\\
        Instituto de Ciencias Nucleares \\
        Universidad Nacional Aut\'{o}noma de M\'{e}xico \\
        A.P. 70-543, C.P. 04510 Distrito Federal, Mexico\\
        E-mail: \email{gerber@correo.nucleares.unam.mx, irais.bautista@correo.nucleares.unam.mx, wolbi@nucleares.unam.mx, he$_{-}$mejia@yahoo.com.mx}}

\author{Christoph P.\ Hofmann \\
        Facultad de Ciencias, Universidad de Colima\\
        Bernal D\'{\i}az del Castillo 340, Colima C.P. 28045, Mexico\\
        E-mail: \email{christoph.peter.hofmann@gmail.com}}

\abstract{In Monte Carlo simulations with a local update algorithm,
the auto-correlation with respect to the topological charge tends to
become very long. In the extreme case one can only perform reliable
measurements within fixed sectors. We investigate approaches to extract
physical information from such topologically frozen simulations.
Recent results in a set of $\sigma$-models and gauge theories are
encouraging. In a suitable regime, the correct value of some 
observable can be evaluated to a good accuracy. 
In addition there are ways to estimate the value of
the topological susceptibility.}

\FullConference{The 32nd International Symposium on Lattice Field Theory,\\
		23-28 June, 2014\\
		Columbia University New York, NY}

\begin{document}

\section{Introduction}

\vspace*{-3mm}
In many relevant models, the configurations are divided into
topological sectors (for periodic boundary conditions). This
includes the $O(N)$ models in $d=(N-1)$, all the 2d $CP(N-1)$ 
models, 2d and 4d Abelian gauge theory, and 4d Yang-Mills theories.
The topology persists if we include fermions,  hence this class 
of models also includes the Schwinger model, QED and QCD.

In the continuum formulation, a continuous deformation
of a configuration (at finite Euclidean action) cannot change the
topological charge $Q \in \Z$. On the lattice there are no
topological sectors in this strict sense, but at fine lattice spacing 
the configurations of the above models occur in distinct 
sectors with local minima, separated by boundary zones of 
higher action. Thus it is possible --- and often useful --- to 
introduce topological sectors also in lattice field theory, although 
the definition of $Q$ is somewhat ambiguous. For the $O(N)$ models that 
we are going to consider, the geometric definition \cite{BergLuscher} 
has the virtue that it naturally provides integer values of $Q$. 

Most simulations in lattice field theory are performed
with local update algorithms, such as the Metropolis algorithm 
for spin models, the heat-bath algorithm for pure gauge theories, 
and the Hybrid Monte Carlo algorithm for QCD with dynamical quarks. 
If there are well-separated topological sectors, such simulations may 
face a severe problem: a Markov chain
hardly  ever changes $Q$. Thus the simulation tends to get stuck 
in one topological sector for an extremely long computation time.
Such a tremendous topological auto-correlation time was observed
{\it e.g.}\ by the JLQCD Collaboration in their QCD simulations with 
dynamical overlap quarks \cite{JLQCD}. For QCD simulations with 
non-chiral quarks ({\it e.g.}\ given by Wilson fermions) the problem
has been less severe so far, {\it i.e.}\ for lattice
spacings $a \gtrsim 0.05~{\rm fm}$ that have typically been used.
However, in the future even finer lattices will be employed,
and then this problem will become manifest.

So how can we measure the expectation value of some observable,
$\la \Omega \ra$, or the topological susceptibility
$
\chi_{\rm t} = 
\frac{1}{V} \left( \la Q^{2} \ra - \la Q \ra ^{2} \right) \ ,
$
if only topologically frozen simulations can be performed?
\footnote{$V$ is the volume, and we will deal 
with parity symmetric models, where $\la Q \ra = 0$.}

L\"{u}scher suggested open boundary conditions, so $Q$ can change 
continuously \cite{openbc}. This overcomes the problem, but giving 
up integer $Q$ has disadvantages, like losing the link to aspects 
of field theory in the continuum, {\it e.g.}\ regarding the
$\epsilon$-regime of QCD.

Here we investigate approaches where periodic boundaries,
and therefore $Q \in \Z$, are preserved. In the framework of
non-linear $\sigma$-models, we test methods to
extract physical results from Markov chains, which
are permanently trapped in a single topological sector,
hence numerical measurements are available only at fixed $Q$.
We start with a procedure to determine $\chi_{\rm t}$ from the
correlation of the topological charge density, which was
introduced by Aoki, Fukaya, Hashimoto and Onogi \cite{AFHO}.
Then we probe a way to assemble an expectation value $\la \Omega \ra$ 
from topologically restricted results $\la \Omega \ra_{|Q|}$. That approach
is based on the Brower-Chandrasekharan-Negele-Wiese (BCNW) formula
\cite{BCNW}, which also yields an estimate for $\chi_{\rm t}$.
\vspace*{-3mm}

\section{Correlation of the topological charge density}

\vspace*{-2mm}
Ref.\ \cite{AFHO} derived an approximate formula for the correlation 
of the topological charge density $q$, at topological charge $\pm Q$
and large separation $|x|$ (we now use lattice units),
\vspace*{-1mm}
\be  \label{denseq}
\la q_{0} \ q_{x} \ra_{|Q|, ~ |x| \gg 1}
\approx - \frac{\chi_{t}}{V} + \frac{Q^{2}}{V^{2}} \ .
\ee
The derivation assumes $\la Q^{2} \ra$ to be large,
and $|Q| / \la Q^{2} \ra$ to be small.\footnote{Actually this 
formula also involves a {\it kurtosis} term (which vanishes for 
Gaussian $Q$-distributions). However, its contribution
is negligible in all examples that we considered,
so here we skip that term.} Therefore we
will limit our considerations to the sectors with $|Q| \leq 2$.
We are going to consider the 1d $O(2)$ model and the 2d $O(3)$
model, and the explicit condition for $\la Q^{2} \ra = \chi_{\rm t} V$ 
will be tested.

In our simulations we use the Wolff cluster algorithm 
\cite{Wolff}, which performs non-local cluster updates. Hence we 
can also measure $\chi_{\rm t}$ directly, which is useful for 
testing this method in view of other models (in particular 
gauge theories), where no efficient cluster algorithm is available.
Preliminary results were anticipated in Ref.\ \cite{Oaxaca}, 
and Ref.\ \cite{chitQCDNf2} presented before a related study 
(with different densities) in 2-flavour QCD.

The 1d $O(2)$ model, or quantum rotor, describes a free quantum 
mechanical scalar particle on the circle $S^{1}$. We use periodic 
boundary conditions in Euclidean time over the size $L$. The continuum
formulation deals with an angle $\vp (x)$, where $\vp (0) = \vp (L)$. 
The lattice variables are the angles $\vp_{x}$, $x = 1, \dots L$, 
with $\vp_{1} = \vp_{L+1}$. We define the nearest site difference as
\vspace*{-1.8mm}
\be
\Delta \vp_{x} = ( \vp_{x+1} - \vp_{x} ) \ {\rm mod} \ 2 \pi
\ \in (-\pi , \pi] \ ,
\vspace*{-1.8mm}
\ee
{\it i.e.}\ the modulus function acts such that it 
minimises  the absolute value. This yields the (geometrically defined)
topological charge density $q_{x}$ and the topological charge $Q$,
\vspace*{-2mm}
\be
q_{x} = \frac{1}{2\pi} \Delta \vp_{x} \ , \quad
Q = \sum_{x=1}^{L} q_{x} \in \Z \ .
\vspace*{-2mm}
\ee

We now give the continuum action and the three lattice actions 
--- standard action, Manton action \cite{Manton} and constraint action 
\cite{constraint} --- that we studied,
\bea
\vspace*{-1mm}
S_{\rm continuum}[\vp ] &=& \frac{\beta}{2} \int_{0}^{L} dx \, \dot \vp(x)^{2} \ , 
\quad S_{\rm standard}[\vp ] = \ \beta \sum_{x=1}^{L} (1 - \cos \Delta \vp_{x}) 
\ , \nn \\
S_{\rm Manton}[\vp ] &=& \frac{\beta}{2} \sum_{x=1}^{L} \Delta \vp_{x}^{2} \ , 
\quad S_{\rm constraint}[\vp ] = \left\{ \begin{array}{ccc}
0 && \Delta \vp_{x} < \delta \quad \forall x \\
+ \infty && \mbox{otherwise} \end{array} \right. \ .
\label{latact1dO2}
\vspace*{-1mm}
\eea
The parameter $\beta$ corresponds here to the moment of inertia, 
and $\delta$ is the constraint angle. The continuum limit
is attained at $\beta \to \infty$ and $\delta \to 0$, respectively.
In the limit $L \to \infty$, $\chi_{\rm t}$ and the correlation length 
$\xi$ are known analytically for all four actions in 
eqs.\ (\ref{latact1dO2}) \cite{rot97,constraint}.
\begin{figure}[h!]
\vspace*{-4mm}
\begin{center}
\hspace*{-2mm}
\includegraphics[width=0.355\textwidth,angle=270]{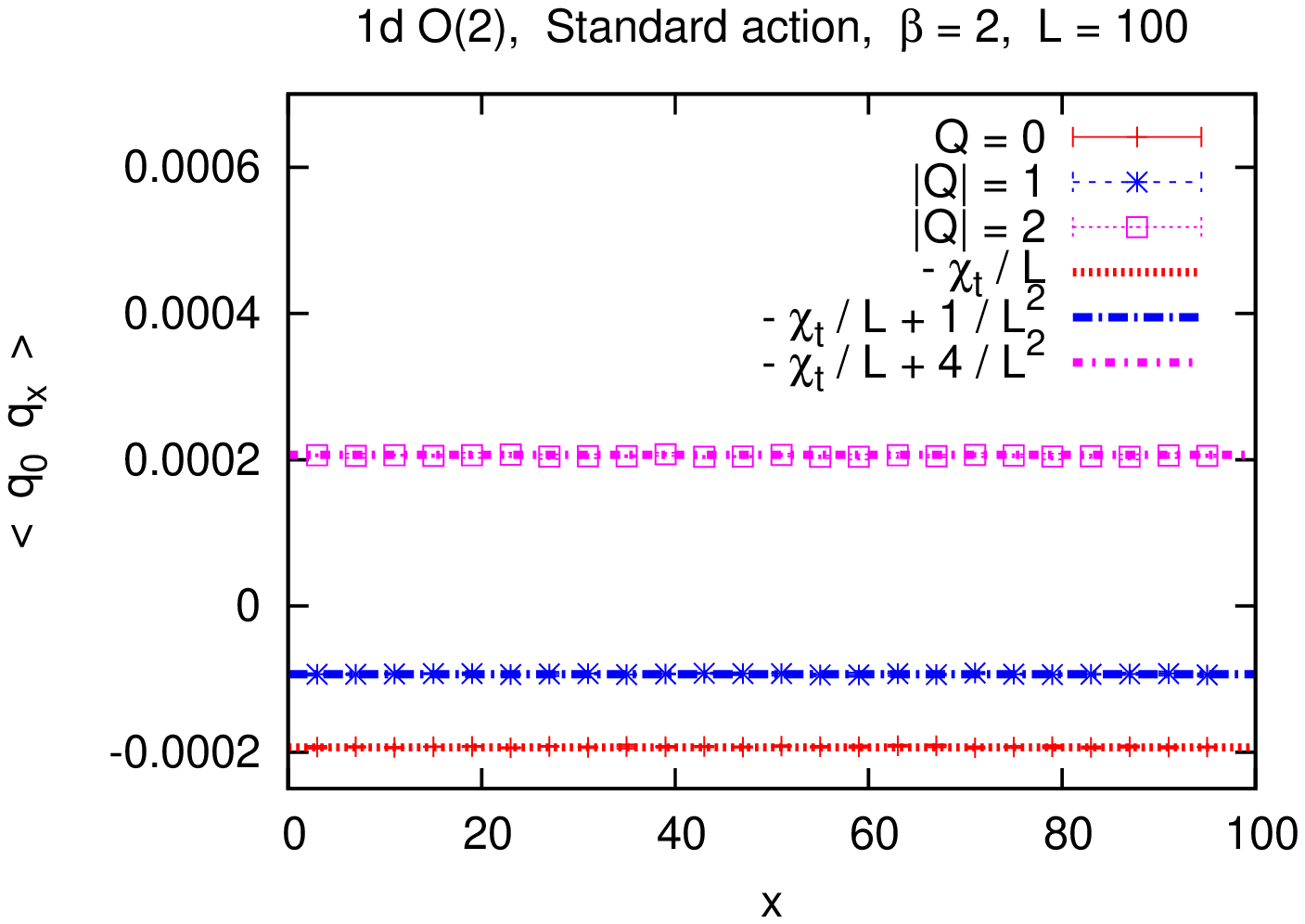}
\hspace*{-4mm}
\includegraphics[width=0.355\textwidth,angle=270]{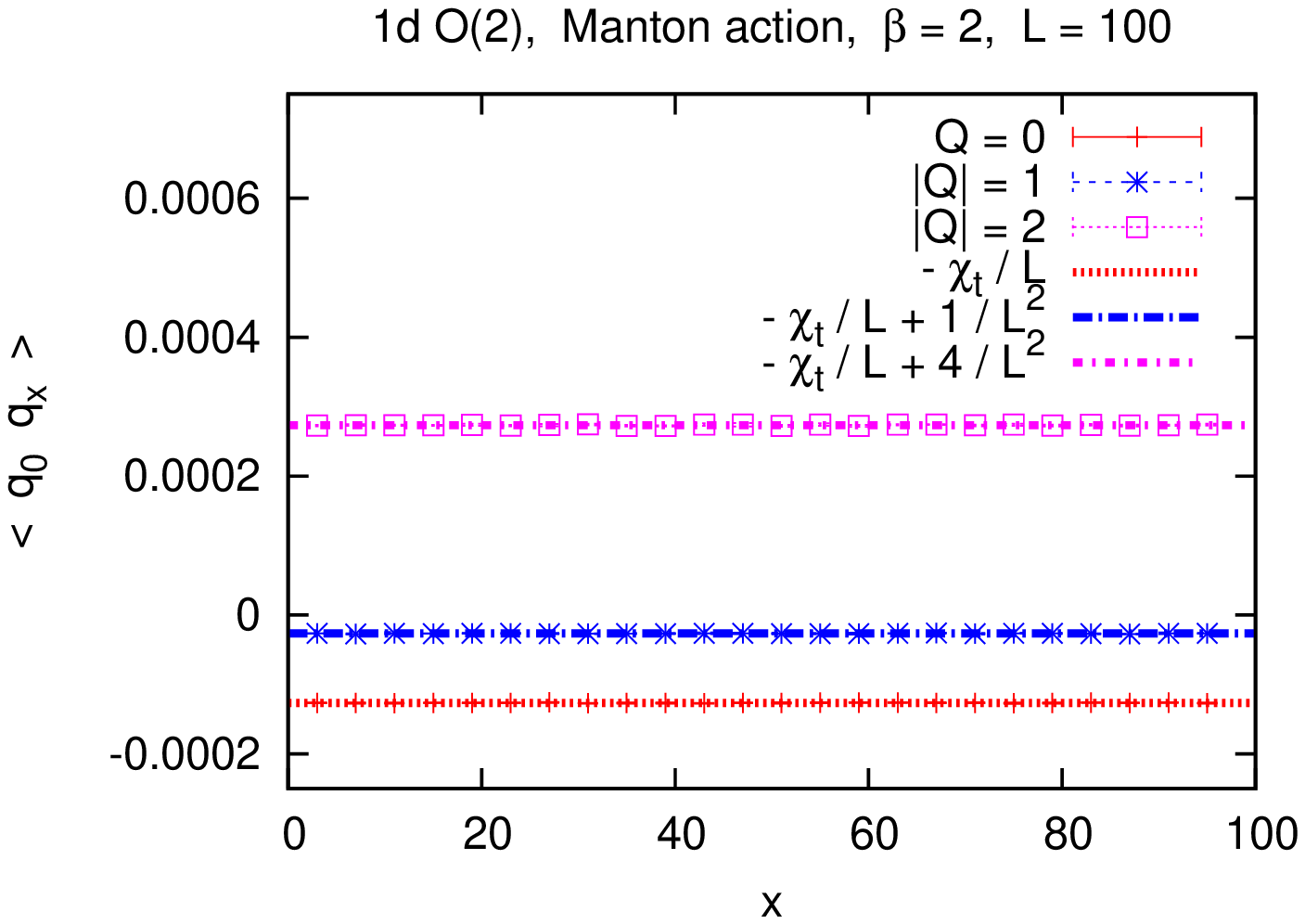} 
\caption{The topological charge density correlation
over a separation of $x$ lattice spacings for the standard action 
(on the left) and for the Manton action (on the right), both at $L=100$
and $\beta =2$. This implies $\xi = 2.779$, $\la Q^{2} \ra = 1.936$
for $S_{\rm standard}$, and $\xi = 4.000$, $\la Q^{2} \ra = 1.266$
for $S_{\rm Manton}$. For comparison, we show the prediction based on 
eq.\ (2.1), where we insert the measured values of $\chi_{\rm t}\,$.}
\vspace*{-7mm}
\label{1dO2SMqq}
\end{center}
\end{figure}

\begin{figure}[h!]
\begin{center}
\hspace*{-2mm}
\includegraphics[width=0.355\textwidth,angle=270]{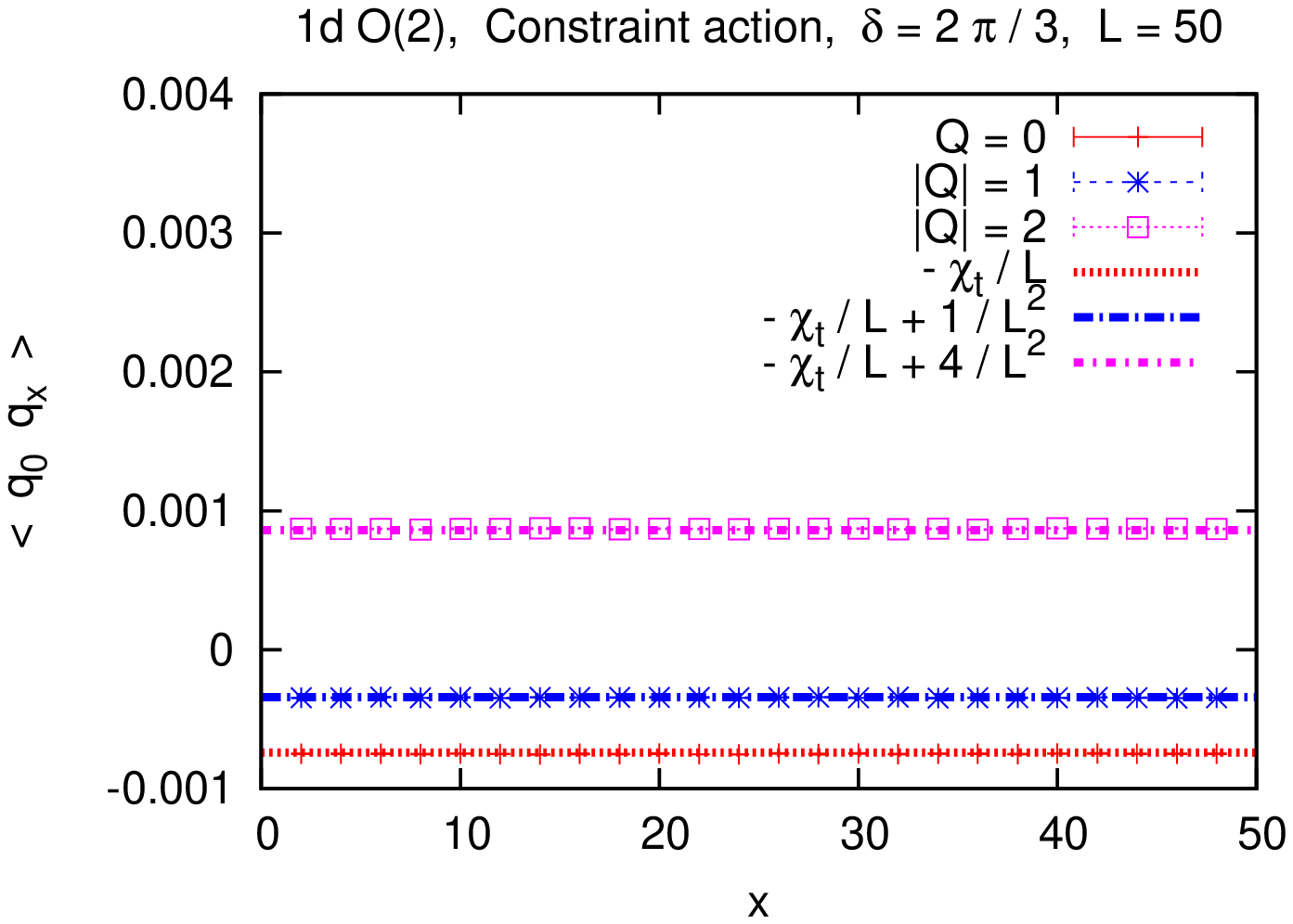}
\hspace*{-4mm}
\includegraphics[width=0.355\textwidth,angle=270]{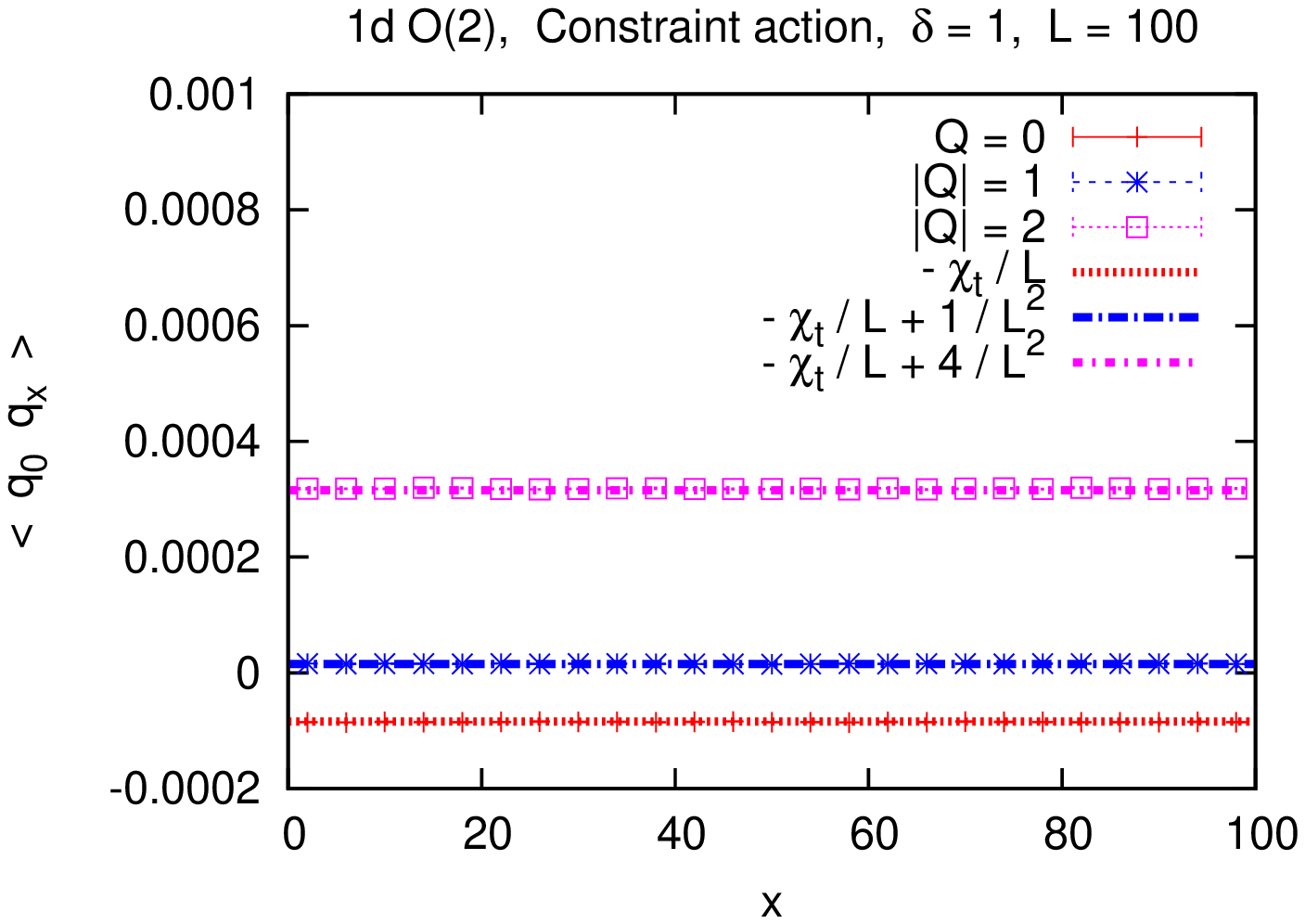}
\caption{The topological charge density correlation over a separation
of $x$ lattice spacings for the constraint action at $\delta = 2 \pi /3$,
$L=50$ (on the left, with $\xi = 1.132$, $\la Q^{2} \ra = 1.852$), 
and  $\delta = 1$, $L=100$ (on the right, with 
$\xi = 5.793$, $\la Q^{2} \ra = 0.844$). 
Again we compare with the prediction (2.1), using the measured
$\chi_{\rm t}$.}
\vspace*{-7mm}
\label{1dO2Cqq}
\end{center}
\end{figure}
Figures \ref{1dO2SMqq} and \ref{1dO2Cqq} show results for the 
standard action, the Manton action and the constraint action at
different sizes $L$ and parameters $\beta$ and $\delta$. 
They are all in excellent agreement with the prediction based on 
eq.\ (\ref{denseq}) (horizontal lines), even down to $\la Q^{2} \ra < 1$.

Next we address the 2d $O(3)$ model, or Heisenberg model, on square 
lattices of size $L \times L$, with classical spins 
$\vec e_{x} \in S^{2}$. Here we simulated the standard action and the 
constraint action, which are analogous to the formulations 
(\ref{latact1dO2}),
\vspace*{-2mm}
\be
S_{\rm standard}[ \vec e ] = \beta \sum_{x,\mu} (1 - \vec e_{x} \cdot
\vec e_{x + \hat \mu}) \ , \quad
S_{\rm constraint}[ \vec e ] = \left\{ \begin{array}{cccc}
0 &&& \vec e_{x} \cdot \vec e_{x + \hat \mu} > \cos \delta
\quad \, \forall x, \, \mu = 1,2 \\
+ \infty &&& {\rm otherwise} \end{array} \right. \ .
\ee
Also here we use the geometric definition of the topological charge,
which is written down explicitly in Ref.\ \cite{constraint}.
\begin{figure}[htb]
\begin{center}
\includegraphics[width=0.38\textwidth,angle=270]{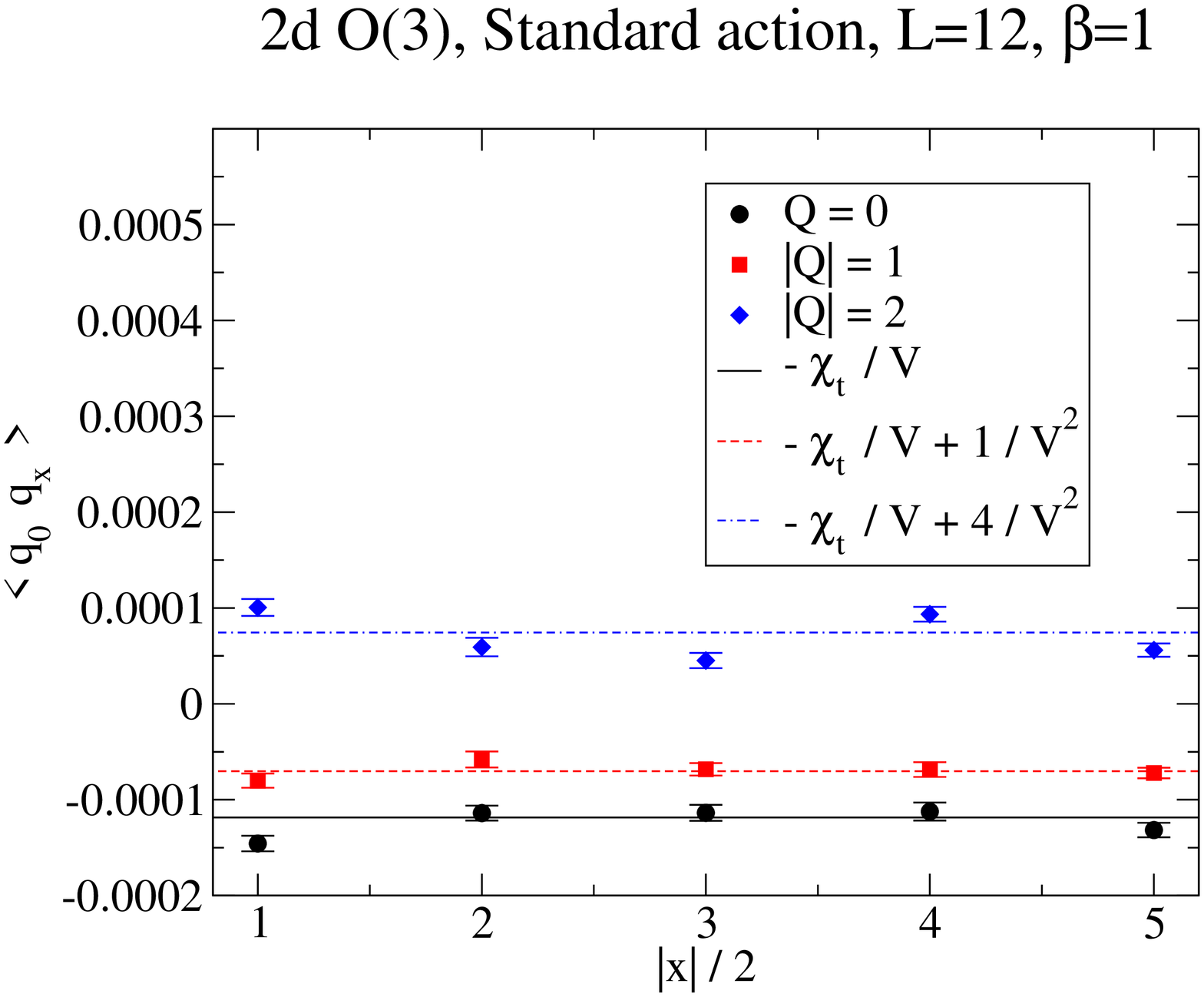}
\hspace*{-6mm}
\includegraphics[width=0.38\textwidth,angle=270]{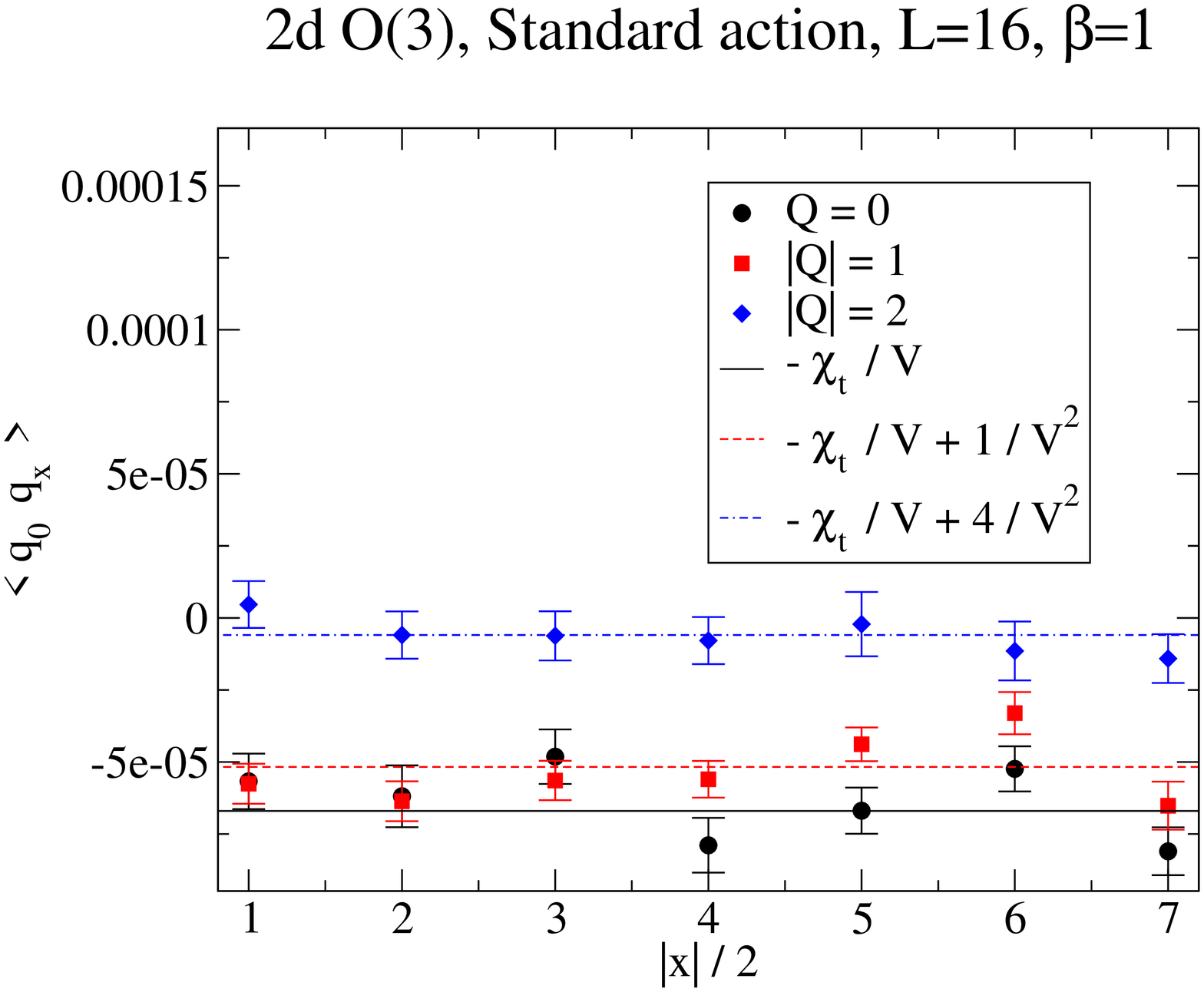}
\vspace*{-2mm} \\
\includegraphics[width=0.38\textwidth,angle=270]{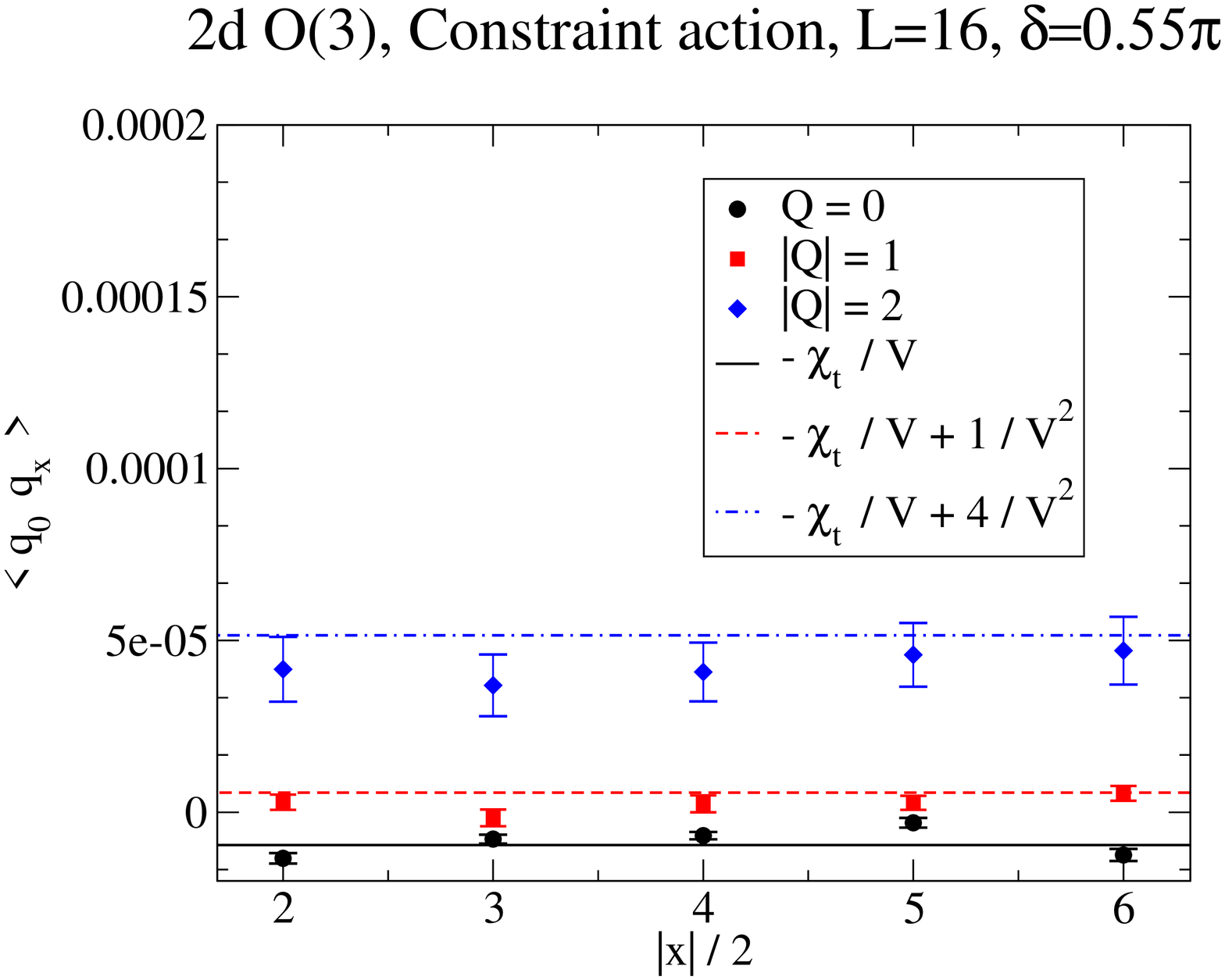}
\hspace*{-6mm}
\includegraphics[width=0.38\textwidth,angle=270]{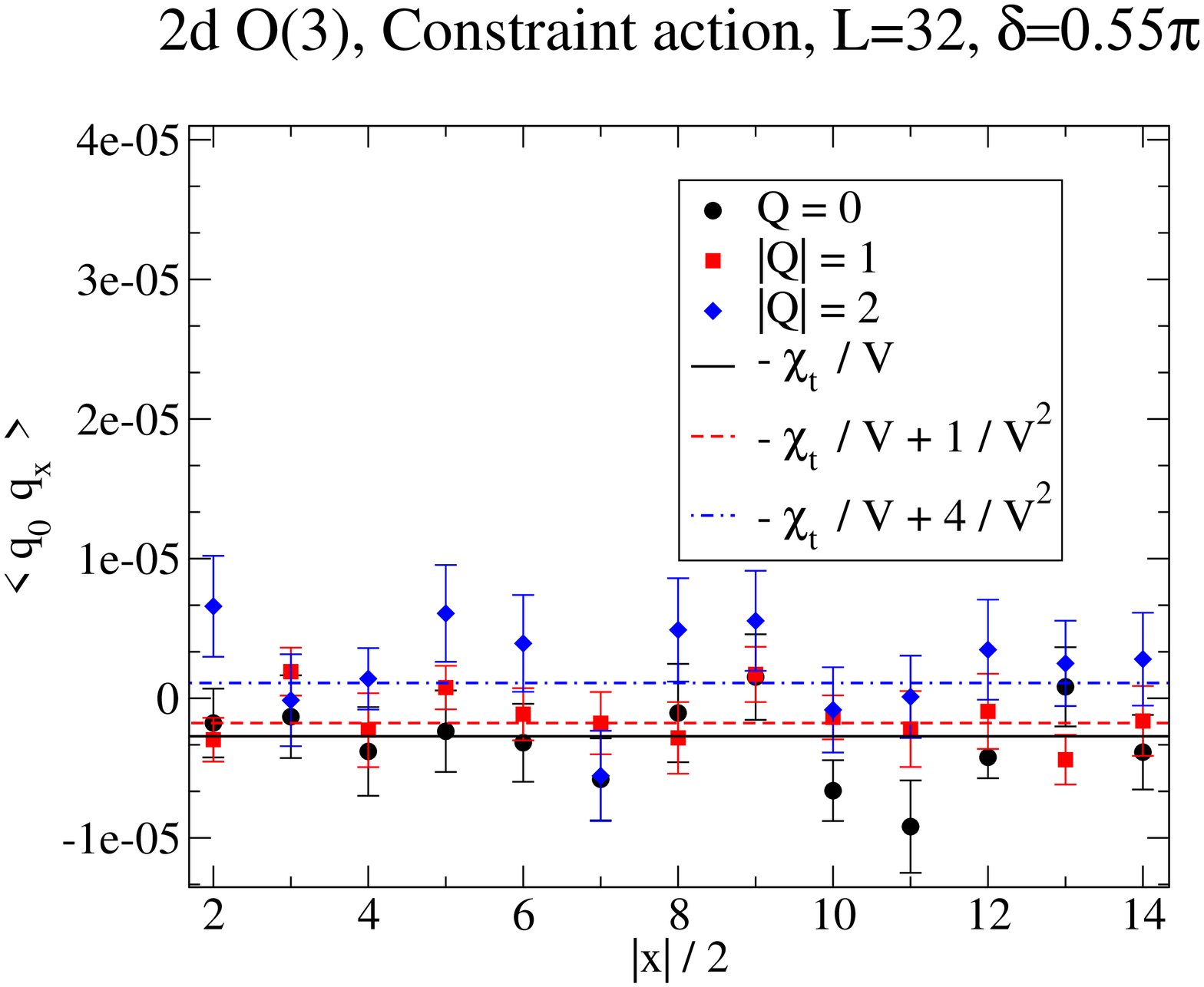}
\caption{The topological charge density correlation in the 2d $O(3)$ model
on $L \times L$ lattices for the standard action at $\beta = 1$ 
(above, $\xi \simeq 1.3$; $\la Q^{2} \ra = 2.46$ at $L=12$ and 
$\la Q^{2} \ra = 4.39$ at $L=16$) 
and for the constraint action at $\delta =  0.55 \pi$ (below, 
$\xi \simeq 3.5$; $\la Q^{2} \ra = 0.63$ at $L=16$ and $\la Q^{2} \ra =
2.86$ at $L=32$). Due to the definition of the topological charge density, 
we proceed in separation steps of 2 lattice spacings. 
In all cases, we show for comparison the prediction based on eq.\ (2.1),
with the measured values of $\chi_{\rm t}$. It works well, even at
$\la Q^{2} \ra = 0.63$ the result is reasonable, although slight
deviations from the prediction show up. But the last plot illustrates 
that for increasing volume the signal get lost in the statistical noise.}
\vspace*{-7mm}
\label{2dO3SCqq}
\end{center}
\end{figure}
Figure \ref{2dO3SCqq} shows results for the topological charge 
density correlation.
Again the comparison to the prediction (\ref{denseq}) works in all
cases (in this context we don't have to worry about the fact that 
$\chi_{\rm t} \cdot \xi^{2}$  diverges logarithmically 
in the continuum limit). However, we also observe that
for increasing volume it becomes soon difficult to resolve
a clear signal from the statistical noise (as required for the
determination of $\chi_{\rm t}$), even with the huge statistics
provided by the cluster algorithm. The quantitative results
will be given in Ref.\ \cite{prep}.

Thus we confirm that formula (\ref{denseq}) is a valid approximation 
over a broad set of parameters. 
Nevertheless, in view of 4d quantum field theory its application
is not promising, since for large volume it becomes very statistics 
demanding. That limitation is in agreement with the conclusion of an
earlier study in the 2-flavour Schwinger model with dynamical overlap 
hypercube fermions (and with the plaquette gauge action) \cite{BHSV}: 
at $\beta =5$, $V = 16 \times 16$ and fermion masses $m= 0.01 \dots
0.06$ a statistics of $O(1000)$ configurations in one topological
sector was insufficient to determine $\chi_{\rm t}$ in this way;
to achieve this to about 2 digits would take at least $O(10^{5})$ 
configurations.

\section{Applications of the BCNW formula}

\vspace*{-1mm}

We now turn to the more ambitious goal of evaluating an observable
$\la \Omega \ra$, when only some values $\la \Omega \ra_{|Q|}$ ---
at various $|Q|$ and volumes --- are available. To this end,
we use an approximate formula, which was derived in Ref.\ \cite{BCNW},
\be  \label{BCNWeq}
\la \Omega \ra_{|Q|} \approx \la \Omega \ra + \frac{c}
{V \chi_{\rm t}} \Big( 1 - \frac{Q^{2}}{V \chi_{\rm t}} \Big) \ .
\ee
Our input are measured values for the left-hand-side in various
$|Q|$ and $V$, and a fit determines $\la \Omega \ra$,
$\chi_{\rm t}$ and $c$, where the former two are of interest.
We refer to a regime of moderate $V$, where these three quantities
practically take their infinite-volume values, but the 
$\la \Omega \ra_{|Q|}$ are still well distinct. 

This is the beginning of an expansion in $1/ \la Q^{2}\ra$, 
hence $\la Q^{2}\ra$ should be large, but what that means has 
to explored numerically. Moreover the assumption of a small value 
of $|Q|/ \la Q^{2}\ra$ is involved again (see also the re-derivation
in Ref.\ \cite{BHSV}), hence we only use sectors with $|Q| \leq 2$. 
An extended expansion has been mentioned in Ref.\ \cite{BCNW} (first
work), and explored in great detail in Refs.\ \cite{Arthur13,CC,Arthur14}. 
It involves further free parameters, and
the crucial question if this improves
the results for $\la \Omega \ra$ and $\chi_{\rm t}$ 
was addressed in Refs.\ \cite{Arthur13,CC,Arthur14}, and will be 
discussed further in Ref.\ \cite{prep}.

\begin{figure}
\vspace*{-4mm}
\center
\hspace*{-2mm}
\includegraphics[width=0.357\textwidth,angle=270]{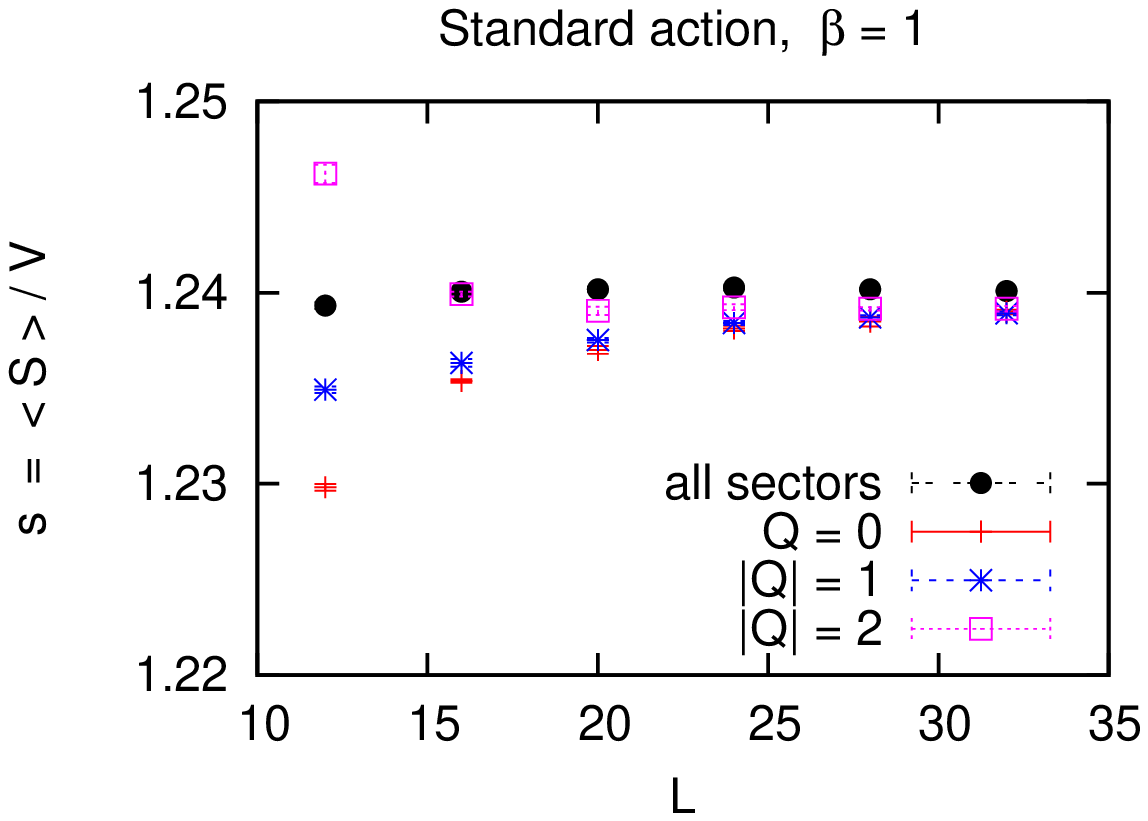}
\hspace*{-5mm}
\includegraphics[width=0.357\textwidth,angle=270]{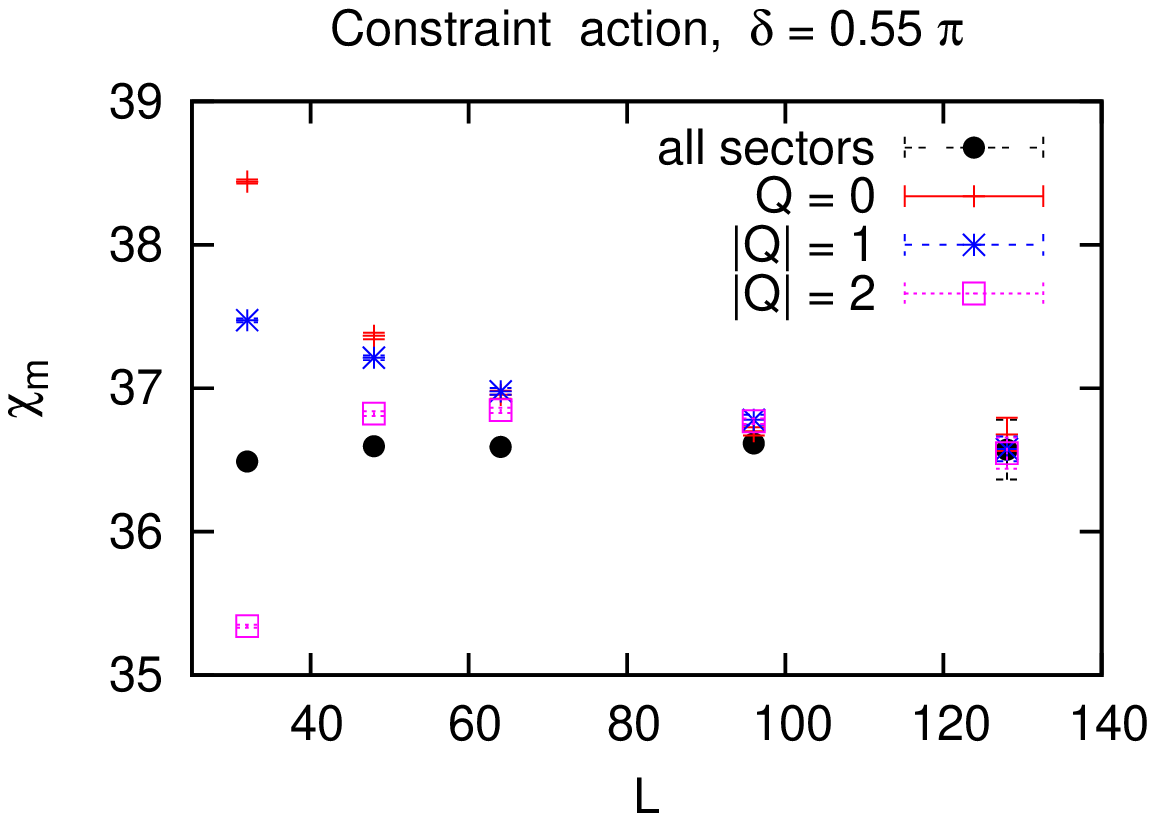} 
\caption{The action density $s = \la S \ra /V$ for the 2d $O(3)$ model
on $L \times L$ lattice (standard action, $\beta =1$, $\xi \simeq 1.3$), 
and the magnetic susceptibility with the constraint action at 
$\delta = 0.55 \pi$ ($\xi \simeq 3.6$). We display the values 
measured in all sectors, and restricted to $|Q|= 0,\, 1$ or $2$.}
\label{magS2dO3fig}
\vspace*{-4mm}
\end{figure}

As our observables we consider the action density
$s = \la S \ra / V$ and the magnetic susceptibility
$\chi_{\rm m} = \la \vec M^{2} \ra / V$ (where $\vec M
= \sum_{x} \vec e_{x}$ is the magnetisation, and 
$\la \vec M \ra = \vec 0$). Results for the 2d $O(3)$ model
in $V = L \times L$ are shown in the plots of Figure \ref{magS2dO3fig}, 
which reveal the aforementioned regimes of ``moderate $V$''. The 
fitting results involving the sectors $|Q| = 0,\, 1, \, 2$, and 
various ranges in $L$ in those regimes, are given in Table 
\ref{magS2dO3tab}. In particular we see an impressive 
precision of the values for $\chi_{\rm m}$, and also the fitting
results for $s$ and $\chi_{\rm t}$ are quite good.

\begin{table}[h!]
\centering
\begin{tabular}{|c||c|c|c||c|}
\hline
{\em Standard action} & & & &  directly measured \\
fitting range for $L$ & $ 16 \ - \ 24$ & $ 16 \ - \ 28$ & 
$ 16 \ - \ 32$ & in all sectors at $L=32$ \\
\hline
\hline
$s $ & 1.24038(12) & 1.24027(8) & 1.24015(5) & 1.24008(5) \\
\hline
$\chi_{\rm t}$ & 0.0173(6) & 0.0169(5) & 0.0164(5) & 0.01721(4) \\
\hline
\hline
{\em Constraint action} & & & &  directly measured \\
fitting range for $L$ & $ 48 \ - \ 64$ & $ 48 \ - \ 96$ & 
$ 48 \ - \ 128$ & in all sectors at $L=128$ \\
\hline
\hline
$\chi_{\rm m} $ & 36.56(4) & 36.58(3) & 36.57(2) & 36.57(2) \\
\hline
$\chi_{\rm t}$ & 0.00262(17) & 0.00256(16) & 0.00259(14) & 0.002790(5) \\
\hline
\end{tabular}
\caption{Above: the action density $s = \la S \ra /V$
extracted from fits to the BCNW formula (3.1), at $|Q|\leq 2$ and
various ranges of the $L$. For $L \geq 16$ the directly measured
$s$ stabilises. It is close to the fitting results. Below:
the susceptibilities $\chi_{\rm m}$ and $\chi_{\rm t}$, extracted from 
fits in various ranges of the $L$. For $L\geq 48$ the directly measured
$\chi_{\rm m}$ stabilises, and the results in distinct sectors converge 
quite well around $L=96 \dots 128$. The fits in moderate volumes
lead to extremely precise values for $\chi_{\rm m}$. 
For both observables, also the fitting results 
for $\chi_{\rm t}$ are correct within less than $2 \sigma$.}
\vspace*{-8mm}
\label{magS2dO3tab}
\end{table}

\section{Conclusions}

In simulations with local update algorithms and fine lattices,
the Monte Carlo history tends to be confined to a single topological
sector for an extremely long (simulation) time. This rises questions
about the ergodicity (even within one sector). Here we do not address this
conceptual issue; we trust the topologically restricted measurements
of $\la \Omega \ra_{|Q|}$, and try to interpret them physically.

In very large volumes $V$, the restricted expectation values all coincide
with the physical result, $\la \Omega \ra_{|Q|} \equiv \la \Omega \ra$,
cf.\ eq.\ (\ref{BCNWeq}), but in practical simulations 
such large volumes are often inaccessible.
For smaller $V$, where $\la \Omega \ra$ 
is well converged to its large-$V$ limit, but the 
$\la \Omega \ra_{|Q|}$ are still significantly distinct, the BCNW 
formula (\ref{BCNWeq}) often allows us to determine $\la \Omega \ra$
to a good accuracy, and it also provides useful results
for $\chi_{\rm t}$. It is favourable to employ only the sectors
with $|Q|\leq 2$, and (roughly speaking) the method is successful
if $\la Q^{2} \ra \gtrsim 1.5$.

If we relax that requirement to $\la Q^{2} \ra \gtrsim 2/3$, 
we can still measure $\chi_{\rm t}$ from the topological 
charge density correlation $\la q_{0} \ q_{x} \ra_{|Q|}$.
The (theoretical) condition of a large separation $|x|$ turns out
to be harmless in practice, but for increasing $V$ the
wanted signal decreases very rapidly. Therefore that method is
hardly promising for 4d models, where --- in reasonable volumes ---
the signal would most likely be overshadowed by statistical noise.

On the other hand, the BCNW formula {\em is} promising for applications
in QCD, where typical simulations take place at $\la Q^{2} \ra = O(10)$. 
This observation is supported by studies in the Schwinger model 
\cite{WBIH,BHSV,CC}, the quantum rotor with a potential \cite{Arthur13} 
and in 4d $SU(2)$ gauge theory \cite{Arthur14}, which will be 
reported in detail in Ref.\ \cite{prep}.\\

\vspace*{-2mm}

\noindent {\bf Acknowledgements:} \
We are indebted to Christopher Czaban, Arthur Dromard, Lilian Prado 
and Marc Wagner for valuable communication and collaboration.
This work was supported by the Mexican {\it Consejo Nacional de Ciencia 
y Tecnolog\'{\i}a} (CONACyT) through project 155905/10 ``F\'{\i}sica 
de Part\'{\i}culas por medio de Simulaciones Num\'{e}ricas'', as well 
as DGAPA-UNAM. The simulations were performed on the cluster of the 
Instituto de Ciencias Nucleares, UNAM.

\end{document}